\documentclass[journal]{IEEEtran}

\usepackage{amssymb}
\usepackage{float}
\usepackage{wrapfig}
\usepackage{multirow}
\usepackage{graphicx}
\usepackage{cite}
\usepackage{tabulary}
\usepackage{amsmath}
\usepackage{graphics}
\usepackage{amsmath}
\begin{document}
\raggedbottom

\title{Who/What is My Teammate? Team Composition Considerations in Human-AI Teaming}

\author{Nathan J. McNeese,
        Beau G. Schelble,
        Lorenzo Barberis Canonico,
        and Mustafa Demir
\thanks{This research was partially funded by NSF Award \#1829008 and AFOSR Award FA9550-20-1-0342 (Program Manager: Laura Steckman).}
\thanks{Nathan J. McNeese and Beau Schelble, are affiliated with the department of Human-Centered Computing in the School of Computing at Clemson University, Clemson, SC 29631. (Email: mcneese@g.clemson.edu; bschelb@g.clemson.edu).}
\thanks{Lorenzo Barberis Canonico is affiliated with the Biomedical Data Science Department at Stanford University, Stanford, CA 94305. (Email: lorenzb@g.clemson.edu).}
\thanks{Mustafa Demir is affiliated with the Ira A. Fulton Schools of Engineering at Arizona State University, Phoenix, AZ 85004. (Email: mdemir@asu.edu).}
\thanks{©2021 IEEE. Personal use of this material is permitted. Permission from IEEE must be obtained for all other uses, in any current or future media, including reprinting/republishing this material for advertising or promotional purposes, creating new collective works, for resale or redistribution to servers or lists, or reuse of any copyrighted component of this work in other works.}}

\markboth{Journal}%
{McNeese \MakeLowercase{\textit{et al.}}: Who/What is My Teammate? Team Composition Considerations in Human-AI Teaming}

\maketitle

\begin{abstract}
There are many unknowns regarding the characteristics and dynamics of human-AI teams, including a lack of understanding of how certain human-human teaming concepts may or may not apply to human-AI teams and how this composition affects team performance. This paper outlines an experimental research study that investigates essential aspects of human-AI teaming such as team performance, team situation awareness, and perceived team cognition in various mixed composition teams (human-only, human-human-AI, human-AI-AI, and AI-only) through a simulated emergency response management scenario. Results indicate dichotomous outcomes regarding perceived team cognition and performance metrics, as perceived team cognition was not predictive of performance. Performance metrics like team situational awareness and team score showed that teams composed of all human participants performed at a lower level than mixed human-AI teams, with the AI-only teams attaining the highest performance. Perceived team cognition was highest in human-only teams, with mixed composition teams reporting perceived team cognition 58\% below the all-human teams. These results inform future mixed teams of the potential performance gains in utilizing mixed teams' over human-only teams in certain applications, while also highlighting mixed teams' adverse effects on perceived team cognition.
\end{abstract}


\section{Introduction}
Teams are a constant in any organization's strategy towards addressing complex and multi-faceted challenges. Decades of research on traditional human-human teams have produced a great deal of knowledge regarding the factors affecting team characteristics, and dynamics \cite{ONeill_Salas_2018}. An essential feature of teaming that has emerged from this literature is the impact of team composition, which refers explicitly to the characteristics and attributes of individual team members \cite{Levine_Moreland_1990}. Team composition is known to play a vital role in affecting team processes, and outcomes such as objective team performance \cite{Spotts_Chelte_2005}, situational awareness \cite{Gorman_Cooke_Pedersen_Winner_Andrews_Amazeen_2006}, and team cognition \cite{Gorman_Cooke_2011}.

Human-artificial intelligence (AI) teaming is a relatively new domain with significant research challenges, receiving increasing attention from team researchers \cite{oneill_human-autonomy_2020}. A human-AI team (HAT) involves one or more intelligent and autonomous agents operating as full-fledged members of a team with human membership \cite{oneill_human-autonomy_2020, mcneese2018teaming}. HATs have received increasing research attention as AI technology has been rapidly advancing \cite{Abadi_2016}. With its continued democratization \cite{schaarschmidt2017tensorforce}, AI will be implemented into teams in much more complex ways than ever before. The influence AI has on team composition will increase, enhancing their effects on team characteristics and dynamics such as team performance, situational awareness, and team cognition. These specific team characteristics and dynamics are of particular interest, as they are each notably affected by the presence of AI agents \cite{demir2017team, mcneese2018teaming, DMC_2020}. Situational awareness and team cognition are also both inextricably tied to team performance \cite{Cooke_Gorman_Myers_Duran_2013, mohammed2010metaphor}, and are essential starting points when researching the effects of composition in HATs.

Team cognition and situational awareness must be sufficiently detailed to understand the difficulties presented by HATs. Team cognition refers to the degree of shared understanding between team members and is often conceptualized through shared mental models \cite{Converse_1993}, which represent "an organized understanding of mental representations of knowledge that is shared by team members" \cite{Mathieu_2005} p.38. Subsequently, the very nature of the construct of team cognition is reliant on factors like implicit and explicit communication \cite{Hanna_Richards_2014}, experience \cite{He_Butler_King_2007}, coordination \cite{cooke2004advances}, and team member perceptions \cite{Resick_Dickson_Mitchelson_Allison_Clark_2010}. Team cognition is therefore known to be subject to changes in team composition through these variables \cite{He_Butler_King_2007, Resick_Dickson_Mitchelson_Allison_Clark_2010}.

Team situational awareness is also subject to the same effects from communication \cite{Sperling_2006}, organizational structure \cite{Sorensen_Stanton_2013}, and changes in team composition \cite{Gorman_Cooke_Pedersen_Winner_Andrews_Amazeen_2006}. This impact is because team situational awareness involves each team member understanding the task situation in a way that allows them to predict and anticipate team outcomes and future task events accurately \cite{nofi2000defining}. To have proper situational awareness, each team member must understand the current situation and then adapt that understanding to the dynamic situation \cite{demir2017team}, making teams with high situational awareness more efficient and effective. For example, if a team manages a town's emergency response resources, they must be aware of every new event as it appears, what that event requires, and what their current responsibilities to the team dictate their action should be.

HATs create challenges for situational awareness and team cognition for various reasons; first and foremost, AI team members in the applied world often do not use natural language processing (NLP) for voice or textual communication due to its unreliability \cite{Chowdhary_2020}. In addition to struggles with NLP, AI struggle to understand things like situational context, facial and hand gestures, and other forms of implicit communication \cite{Pereira_Prada_Paiva_2012}. Humans can understand higher dimensions of communication such as context, intent, and non-verbal communication and rely on them heavily to relay information \cite{Fiore_Salas_Cuevas_Bowers_2003}. This contrast in human and AI behavior means team communication patterns, information processing, and interaction strategies exist in a dichotomous state within HATs. Subsequently, when this contrast in team characteristics and dynamics is coupled with human's potential bias towards AI \cite{Demir_perception_2018}, significant adverse effects to team cognition and situational awareness occur, which are detrimental to team outcomes like performance and satisfaction.

Understanding the effects that team composition may have on HATs will allow AI researchers and developers to create more informed AI for multiple settings. This knowledge would also inform HAT practitioners of a potential need to train stakeholders with the autonomous agent(s) they will be working with to help reduce potential bias and even inform interface design for HATs to enhance team cognition and situational awareness. In order to accomplish such research, the following experiment analyzes three types of teams (human teams, AI teams, and HATs) from both a quantitative perspective (looking at the comparative and subjective performance of each team type's outcomes) and a descriptive perspective (using teamwork surveys to quantify the perception of team cognition within their team).

The following research questions serve as the main objectives of this experiment and answer the following research goal:

\textit{Investigating the effects of team composition on human-AI team characteristics and dynamics}
\begin{itemize}
    \setlength{\itemindent}{4em}
    \item [RQ1:] How are team characteristics and dynamics similar and or different across human, human-AI, and AI teams?
    \begin{itemize}
    \setlength{\itemindent}{4em}
        \item [RQ1.1:] In regard to perceived team cognition? 
        \item [RQ1.2:] In regard to team situational awareness?
        \item [RQ1.3:] In regard to team performances?
    \end{itemize}
    \item [RQ2:] How can these similarities and differences be utilized to inform future human-AI teams?
\end{itemize}

\section{Background}
The current paper addresses the effects of team composition on team characteristics and dynamics. In doing so, the relevant literature leading to such questions is reviewed—specifically, the role of communication in teaming and the transition into human-AI teamwork.

\subsection{Role of Communication and Composition Within Teams}

Communication is a foundational aspect of teaming and supports coordination towards a common objective. Variations in team communication strategies across interactions in tasks and sub-tasks are a known predictor of performance \cite{mcneese2014integrative}. Researchers have taken to characterizing teams as information processing units that allow for the attention, encoding, storage, retrieval, and processing of information \cite{hinsz1997emerging}. This process contributes to creating team characteristics and dynamics such as team cognition, team situation awareness, and performance. HATs have also been shown to benefit from voice-based communication, even in high stress/risk contexts like medical first responders' training that was being supplemented by a synthetic agent \cite{Damacharla_Dhakal_Stumbo_Javaid_Ganapathy_Malek_Hodge_Devabhaktuni_2019}. Ideal communication within HATs would generally center around NLP for verbal communication or textual communication \cite{wang_slack_2019}; however, NLP is not the only form of communication in teams. Non-verbal communication also plays a vital role in teamwork, especially in teams that are physically collocated \cite{tiferes_are_2019}. Non-verbal communication consists of eye movements, facial expressions, and hand gestures, which can be just as necessary within teams as verbal or textual communication \cite{warkentin_training_1999}. Despite having strength in collocated teams, non-verbal communication also exists within digital teams in the form of images, emojis, and even punctuation \cite{warkentin_training_1999}.

Communication is another vital component of team cognition and situational awareness development and usage, both of which are significant predictors of team outcomes. Team cognition develops a shared understanding of teamwork and taskwork among team members \cite{mohammed2010metaphor}. Team cognition is both a process (communication and coordination) and an output (shared mental model) \cite{fiore2010toward}, and that output itself then supports team performance. Team cognition is essential to teaming based on research that has directly linked it to team performance, as researchers have noted on multiple occasions that a lack of or breakdown in team cognition may lead to decreased team performance \cite{wilson2007errors}. Team situational awareness is another emergent team dynamic capable of predicting team performance \cite{bolstad2005modeling} and is affected by changes in communication and interaction patterns. As previously stated, team situational awareness involves team members perceiving their dynamic situation in the same way and involves perception, comprehension, and prediction towards future events the team may face \cite{nofi2000defining}. Unfortunately, AI in HATs have great difficulty addressing either type of communication due to unreliable NLP technology \cite{Chowdhary_2020}, though attempts have been made to address this in mixed composition HATs with frameworks like KAoS HART services. The framework was specifically designed to implement more natural and effective communication to help HATs produce better team cognition, and situational awareness \cite{Bradshaw_2009}.

Communication becomes increasingly interesting when considering its dynamic relationship with team composition. The role of team composition in team dynamics is impressive in that it is capable of affecting objective team performance \cite{Spotts_Chelte_2005}, situational awareness \cite{Gorman_Cooke_Pedersen_Winner_Andrews_Amazeen_2006}, and team cognition \cite{Gorman_Cooke_2011}. As teams grow in numbers, communication and interaction patterns are changed based on the growing communication networks within the team. Further, the bandwidth provided to other individual teammates is reduced; this concept is known as the span of control and continues to be relevant in modern teamwork research \cite{el-khalil_managing_2016}. For instance, communication can be observed in teams of varying sizes within a similar environment to understand the optimal team size regarding a team's span of control to create high-quality communication \cite{holm-petersen_size_2017}. It is unknown how the span of control applies to HATs as AI teammates generally do not support the extensive communication necessary to reduce communication bandwidth in teams. Additionally, team composition can also change based on the tasks that teams are responsible for completing. For example, some teams may be held back in tasks that do not allow for all available implicit communication (i.e., virtual team members) \cite{Fiore_Salas_Cuevas_Bowers_2003}. Tasks can be further broken down using Steiner's taxonomy, which characterized and categorized task environments to create an understanding of how team dynamics differ between different tasks; examples of Steiner's tasks include additive, optimizing, and divisible \cite{steiner1972group}. The current research study focuses explicitly on optimizing tasks, where team dynamics interact to create an optimal solution to the task.

Team composition in HATs significantly affects communication and interaction patterns, which may change how team dynamics like situational awareness and team cognition manifest and display themselves. Studies have shown that artificial agents in HATs disrupt the regular communication of information from their lack of human-like behavior \cite{Demir_McNeese_Cooke_2016} and create rigid coordination systems \cite{Demir_Likens_Cooke_Amazeen_McNeese_2018}. These are harmful effects that reduce adaptive behavior due to poor situation awareness and team cognition and create adverse teaming environments that lead to ineffective teams. Understanding these shortcomings requires further research on how varied HAT compositions exhibit team dynamics like situational awareness and team cognition and how they relate to team outcomes. Specifically by improving team outcomes through Bradshaw and colleagues' criterion for effective coordination in HATs, which were inter-predictability (situational awareness), common ground (team cognition), and directability \cite{Bradshaw_Feltovich_Johnson_2011}. These requirements map directly onto the variables of interest in the current study, which will clarify the effects of HATs' dynamic communication and interaction patterns to achieve more effective coordination.

\subsection{Towards Human-AI Teamwork}

The modern field of human-AI interaction has roots in Paul Fitt's early work on function allocation "men are better at/machines are better at" (MABA-MABA approach) \cite{Fitts_1951}. Function allocation was subsequently phased out to move the field towards a more collaborative approach between humans and agents \cite{Bradshaw_Dignum_Jonker_Sierhuis_2012}. Until recently, most research was restricted to the robotic domain or contexts where the human had a supervisory role over an automated system \cite{gao2016designing}. However, breakthroughs in cognitive modeling and AI recently opened up the opportunity to study humans interacting with AI at a peer-level in other teaming structures \cite{sun2006cognition}. For example, research on shared control where humans and robots work together in real-time on a single system has benefited, such as when a surgeon works with a robot to perform surgery \cite{abbink_topology_2018}. Another human-agent collaboration structure that has benefited from a shift to autonomy is mixed-initiative, where humans handle analytical work in data sets filtered and recommended by an AI teammate \cite{hu_dive_2018}, drastically increasing the size of the environment and data sets they analyze to great benefit \cite{chan_solvent_2018,hussein_mixed_2018}.

Researchers have examined this opportunity for advanced HAT research and highlighted major challenges facing constructs like team cognition and situational awareness \cite{Klien_Woods_Bradshaw_Hoffman_Feltovich_2004}. Studies on HATs have shown promise for situational awareness \cite{demir2017team}, team cognition \cite{DMC_2020}, and team coordination and performance \cite{mcneese2018teaming}. The increasing usage of AI has also opened the door to artificial teammates being autonomous instead of automated. This distinction is important as the primary defining factor between the two is how autonomous agents act intelligently to decide their course of action through independence, self-governance, and proactivity \cite{oneill_human-autonomy_2020}. On the other hand, automation is limited by its inability to independently engage in activities that benefit the team without being pre-programmed. Automated agents are also not considered full team members for these reasons, while autonomous agents are \cite{mcneese2018teaming}. These AI agents are also differentiated from expert systems by their ability to learn tasks without being pre-programmed to do so, while expert systems are handcrafted meticulously to operate over a very narrow task space \cite{Phulera_2017}.

Whether the human-human teamwork paradigm's assumptions will end up translating to HATs is an open question. This questioning is caused by the issues posed by NLP's unreliability \cite{Chowdhary_2020}, AI's inability to understand any non-verbal (implicit) communication \cite{Pereira_Prada_Paiva_2012} and differences in interaction patterns between humans and AI due to its expert-level ability \cite{foerster2018learning}. Accordingly, the prospect of HATs developing team characteristics and dynamics similar to what is seen in typical human-human teams seems suspect. However, this is not to say that such dynamics have not been investigated or examined in HATs. A review of the HAT literature has identified several vital metrics common to HAT research \cite{Damacharla_Javaid_Gallimore_Devabhaktuni_2018}. Standard team level metrics include team effectiveness, cohesion, and human-robot ratio, which together investigate HAT characteristics and dynamics to ascertain how they produce their team outputs. Similarly, this review's standard human metrics include shared mental models (an operationalization of team cognition) and situational awareness, which are both metrics analyzed in the current study. This HAT research found that NLP and team cognition lead to higher performance \cite{DMC_2020}, and HAT situation awareness benefits from information pushing \cite{Demir_McNeese_Cooke_2017}.

Communication is vital to team cognition, performance, and team situational awareness, different compositions of humans and AIs will impact team cognition and situational awareness. This assertion has been noted in past human-robot teaming taxonomies \cite{yanco2004classifying}. These taxonomies state that different team compositions and skill levels can pose challenges in terms of how the robots help other team members accomplish the goal, effectively changing interaction patterns \cite{tang2006peer}. In light of the definition of autonomy and HATs outlined previously, research specifically targeting team composition is lacking in the human-AI teaming literature outside of Bradshaw and other's work on the KAoS HART services framework \cite{Bradshaw_2009}. As outlined above, team composition and interaction patterns will affect team dynamics like situational awareness and team cognition, and these effects are necessary to understand how to create effective HATs. Accordingly, the present research will address this gap and shed light on these effects to implement better HATs.

\section{Methods}

\subsection{Participants}
All relevant approvals for research with human subjects were obtained from the Clemson University Institutional Review Board before conducting research. 60 participants (Age: \textit{M} = 30.67, \textit{SD} = 8.16) were recruited for this study from Amazon Mechanical Turk (MTurk) to play NeoCITIES (described in the following subsection) under three different conditions: human-only team, human-human-AI team, human-AI-AI team. Each condition consisted of 10 teams for 30, 20, and 10 human participants per condition, respectively (AI-only condition also consisted of 10 teams). Participants were randomly assigned to one of the three human conditions. Participants spent 40 minutes completing the experiment and were compensated \$3 for their time.

MTurk has been known to possess validity-related advantages compared to traditional lab studies \cite{paolacci2010running}. Utilizing Mturk offers a more diverse pool of subjects from which to draw samples, and the platform's user base is closer to the US population as a whole than university subjects \cite{paolacci2010running}.

\subsection{NeoCITIES Task and Roles}
The team simulation used in this study was NeoCITIES, which recently underwent a rigorous redesign that allowed the integration of AI team members \cite{Schelble_in}. The NeoCITIES experimental platform has a long history in team research and has been used for a variety of previously published team research \cite{klein2009supporting,pfaff2012negative,mohammed2015time}. The redesign included a map for human participants, a complete UI overhaul (seen in Figure 1), and a back-end architecture redesign that allowed for simultaneous sessions and real-time game state tracking (for more details on the NeoCITIES redesign, please see \cite{Schelble_in}). The NeoCTIES task simulation requires three separate team members to coordinate and complete a complex task, which develops various constructs essential to effective team outputs like situational awareness \cite{nofi2000defining} and team cognition \cite{Converse_1993}. The simulated task is an emergency response management scenario within a fictional city that requires three individual players to assume three unique roles (1) Hazmat response, (2) Police response, (3) Fire response. Each player was responsible for managing a specific group of resources for the city (e.g., bomb disposal, SWAT team, patrol car, ambulance, fire truck). More details on specific resources available to players are shown in Table \ref{Role_Resources}.

\begin{table}[h]
    \centering
    \caption{NeoCITIES Role Resources}
    \label{Role_Resources}
            \resizebox{\columnwidth}{!}{%
            \begin{tabular}{|c|c|}
                \hline
                 \textbf{Role} & \textbf{Resources} \\
                 \hline
                 Hazmat Response & Investigator, Bomb Squad, Chemical Truck \\
                 \hline
                 Police Response & Investigator, Squad Car, SWAT Van \\
                 \hline 
                 Fire Response & Investigator, Ambulance, Fire Truck \\
                 \hline
            \end{tabular}%
            }
\end{table}

The resources began at an individual player's home station and were dispatched to events and or recalled back to the home station. The players had to work together to allocate their scarce resources to events around the city's map, with each event requiring specific resources from different players, often in a specific order. A sample of events and their required resources are shown in Table \ref{Event_Resources}. The NeoCITIES simulation also accounted for distance to events, as each resource had a specific speed with which it could travel around the city, making time and distance additional considerations for the team. Slower resources included the chemical truck, ambulance, and investigators. The fastest resource was the fire truck, with all other resources being equal. Once a resource arrived at its destination, it immediately completed its job with no procedural tasks necessary to complete upon arrival. Once reaching the destination, the resource could be brought back to the home station or tasked elsewhere. Events in this experiment were programmed to activate at specific time points throughout the task, with some overlap to increase task difficulty. Additionally, the tasks had to be completed within a set time, or they expired, and the team lost out on potential points towards their final score.

\begin{table}[h]
    \centering
    \caption{Sample NeoCITIES Event's and Required Resources}
        \label{Event_Resources}
            \resizebox{\columnwidth}{!}{%
            \begin{tabular}{|c|c|}
                \hline
                 \textbf{Event} & \textbf{Required Resources} \\
                 \hline
                 Football Weekend Briefing & Investigator \\
                 \hline
                 Tanker Collision & Squad Car, Fire Truck, Chemical Truck \\
                 \hline 
                 Escort a Senator & SWAT Van \\
                 \hline
                 Smoking Kills & Fire Truck \\
                 \hline
                 Field Chemical Removal & Chemical Truck \\
                 \hline
                 Luncheon Nausea & Ambulance, Investigator \\
                 \hline
                 Possible Student Rave & Investigator, Squad Car \\
                 \hline
                 Old Main Frame Shoppe Fire & Investigator, Fire Truck \\
                 \hline
            \end{tabular}%
            }
\end{table}

\begin{figure} [h!]
    \label{NeoCITIES_Home}
  \centering
    \includegraphics[width=0.3\textwidth]{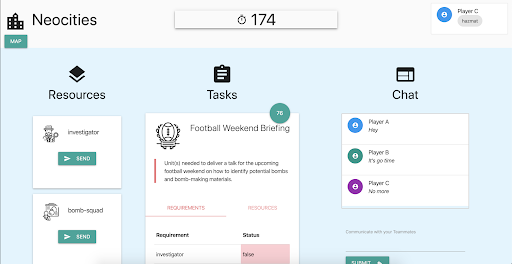}
    \caption{Interface of NeoCITIES}
\end{figure}

NeoCITIES also implemented a chat component, seen in the far right of Figure 1. This chat feature identifies each player and their messages through distinct colors. Chat was only available to human participants as AI agents did not communicate with their team. The resources available were displayed in the far left of Figure 1, and their role was displayed in the top right. There was also an active event component of the UI, seen in detail in the center of Figure 1. Each event was represented by a card containing an event description alongside its current status, broken down by requirement (false/red means that particular resource has not arrived at the event yet). A timer conveying how much time was left before the event failed was included as well, and if the RESOURCES tab was clicked upon, the card then showed the amount of time it would take for each of the player's resources to reach the event. 

Lastly, NeoCITIES implemented a map as an optional toggle (seen in the top left of Figure 1) to enable human users to locate events and resources in real-time. The events and resources were represented through the same icons used on the default view and identified through an available legend. For replicative purposes, the map is vital as AI team members receive a matrix containing geospatial information. Therefore, their human teammates must also receive this information to allow for effective comparisons to be drawn.

The NeoCITIES backend utilized a cloud-based design where clients push updates to a cloud server (Firebase), automatically updating a global state that each client is listening to through a REST API. The Firebase server maintained a JSON tree to store the continuously updated data as the session was being played. A single iteration of the NeoCITIES game took an average of five minutes (five iterations per team for 25 minutes total) and began once all players signed on. At the end of the iteration, relative scores were computed and displayed to the players.

\subsection{AI Agents and Training}
This experiment necessitated the training and creation of AI agents to complete the NeoCITIES task simulation as a team. The AI utilized in this experiment were created using reinforcement learning (RL), a machine learning model that learns tasks through behavioral rewards and has recently gained attention for its superhuman performance in Go, Chess, soccer, and Atari games \cite{foerster2018learning}. Using RL enables NeoCITIES to incorporate the most cutting-edge AI models and enables the agents to generalize to a wide variety of scenarios since RL agents learn independently from the game's feedback.

The agents created for this experiment utilized Proximal Policy Optimization (PPO) \cite{schulman2017proximal} through the Tensorforce RL library. Tensorforce is an open-source RL library focused on providing clear APIs, readability, and modularization to deploy RL solutions in both research and real-world applications \cite{schaarschmidt2017tensorforce}. The only custom coding necessary to create the RL agents was to recreate the NeoCITIES simulation programmatically so the agent could learn through self-play as part of a multi-agent system that encompassed all three roles. Specifically, the system represented NeoCITIES as a sequence of matrices, updating each turn to match the game's ever-evolving state. The agent received the state as a 50x50 matrix that contained the information necessary to complete the simulation successfully and learn over time. The state included identifiers for locating resources, a coordinate-based map of the town, and events with their coordinate location. After receiving the state, the agent acted by selecting each resource's destination coordinates. The environment processed the action, locations of the resources, the status of the events changes, and updated the state. A reward was computed from the updated state based on how much closer the resource was to an appropriate event than its prior state, and the agent then received the reward. The agent's model was produced over 200,000 iterations of this simulation. A single AI model was trained to play all three roles at once, allowing that model to play any role while retaining the knowledge on the needs and responsibilities of the others. As a result of being trained to dispatch resources quickly to the appropriate events, the AI teammate behaved methodically as it was trained to maximize its reward. This training meant the AI would dispatch the appropriate resources to events as they came, similar to any expert-level human, to maximize the possible score. As such, the human team members were met with an AI that was explainable and easy to understand, as it merely cooperated and interacted with its human teammates most appropriately to maximize the team's score.

Lastly, participants interacted with the AI through an API built to enable the trained agent to interact with the Firebase database. This feature allowed the AI agent to act as a teammate to the human participants with no additional workload or interaction from the participant. The AI acted as a full member of the team as any other human would, except for communicating textually. The API connected to Firebase and generated a game state every time there was an update and polled the AI model for an action, which was uploaded to Firebase and reflected in all team member's simulations. The agent and interaction architecture is displayed in Figure 2.

\begin{figure} [h!]
    \label{arcitecture}
        \centering
        \includegraphics[width=0.18\textwidth]{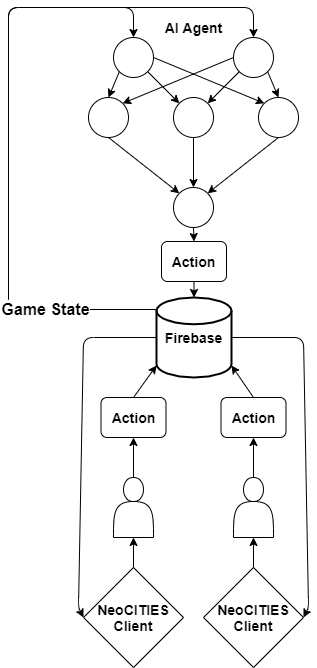}
        \caption{Agent Interaction Architecture}
\end{figure}

\subsection{Procedure}

Upon recruitment from Amazon Mechanical Turk, each participant logged into NeoCITIES through a unique URL. The URL identified the team they were to be a part of, the session they were about to start, and their assigned role (Hazmat, Fire, Police). Participants were randomly assigned to one of the 3 NeoCITIES simulation roles, while AI agents always participated as the Hazmat Response role in the human-human-AI condition. AI agents operated as the Fire and Hazmat Response roles in the human-AI-AI condition. These roles were fixed in order to ensure the manipulation remained constant across teams. Additionally, the constant role of the AI would not have had any effect on the results as each role held equal responsibility to the team and their collective efforts. Participants were made aware upon recruitment on Mechanical Turk that they would be completing the simulation with AI teammates if their condition included such agents. Additionally, participants were informed what team roles their AI teammates would take on during the simulation. Once participants logged in and completed the informed consent form, they were taken to the training page for the NeoCITIES simulation. The training page includes a text explanation of how NeoCITIES functions and how to interact with it. The training page accompanies the text descriptions with short videos showcasing the various actions of NeoCITIES like dispatching resources, using the map, and where events are displayed, as seen in Figure 3. Participants were unable to continue until all participants had viewed and scrolled to the bottom of the training page.

\begin{figure} [h!]
  \centering
    \label{Neo_Training}
    \includegraphics[width=0.3\textwidth]{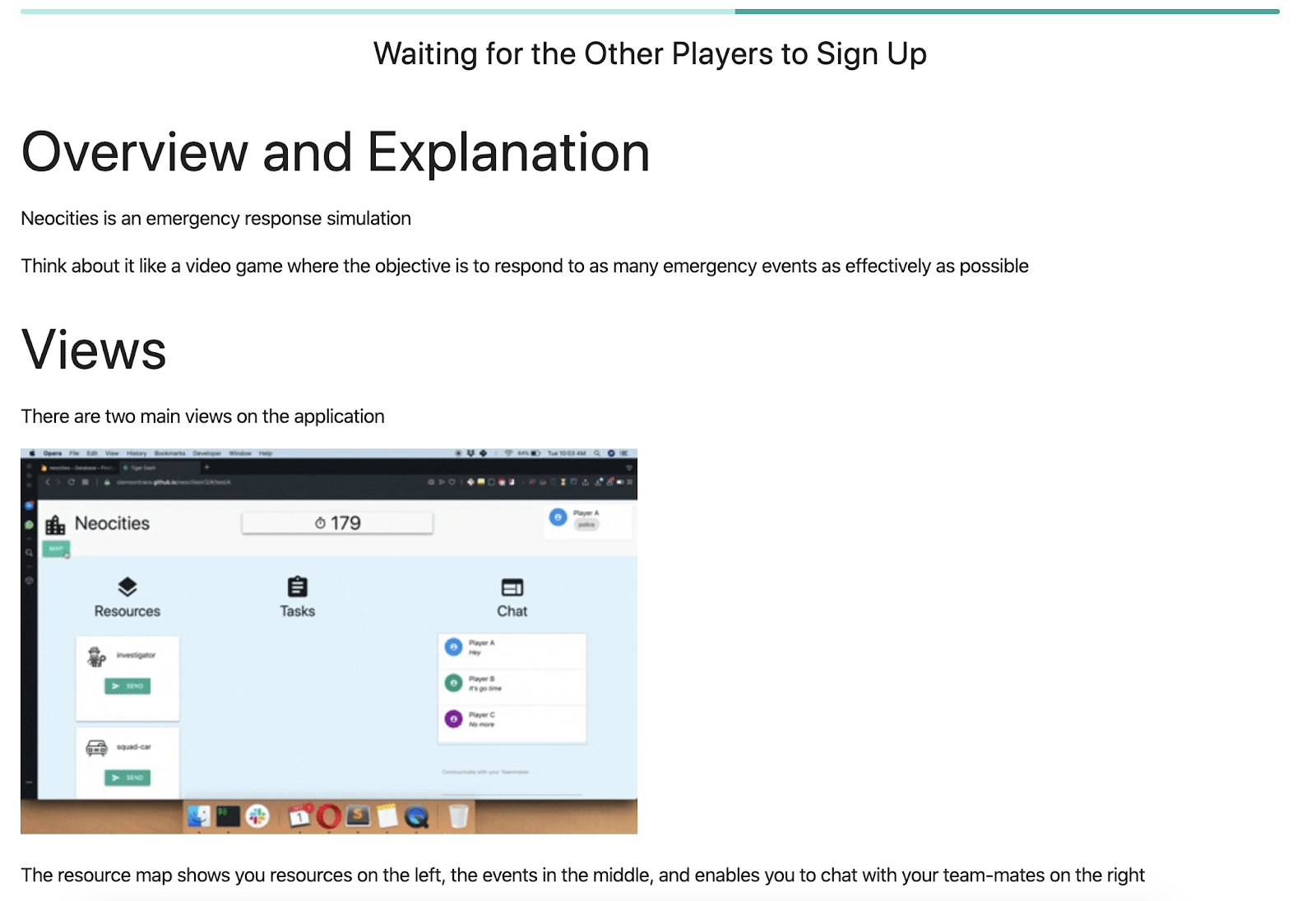}
    \caption{Training Page}
\end{figure}

Once all players were logged in, the interface took the players to the main view, and the session began. The order of emergencies was the same for each team in each condition to ensure consistency across conditions and teams for later comparison. After five minutes elapsed, the relative score was automatically computed and displayed to each player, and a link at the bottom took them to the next session. Once participants completed five sessions, they were taken to Qualtrics to complete a survey that measured their perceptions of team cognition within their team. The total length of time spent completing the five sessions of NeoCITIES was 25 minutes, while the final survey took 15 minutes for a total time of 40 minutes. For the AI-only team condition, the model played five iterations of the game 10 times to create 10 teams.

\subsection{Measurements}
\subsubsection{Team Performance}
NeoCITIES tracks performance by evaluating the speed and accuracy of the team's response through raw (cumulative) and relative (weighted) scores. These scores are based on validated and historical scoring equations that are embedded in NeoCITIES \cite{hellar2010neocities}. Teams are rewarded for rapidly sending the correct type, and amount of resources to an emergency event but are penalized when an event terminates either because of the team's inaction or incorrect and slow response \cite{hellar2010neocities}. The validated and historical scoring equation evaluates team task performance, including AI and human team members' performance. The current formula is designed to incentivize participants to respond to events correctly and timely by allocating resources. Formula (1) and (2), which are variations of the 2010 equation (see \cite{hellar2010neocities}), does not implement random seed coefficients in order to maximize replicability:

\begin{equation}
    Event Score = \frac{(end - start)}
    {(limit - start)) * difficulty}
\end{equation}

\begin{equation}
    Team Score = \frac{100 * [(worstScore - rawScore)}
    {(worstScore - bestScore + 1)]}
\end{equation}

Each event had a difficulty level ranging from one to three, which depended on the event's actual difficulty. This difficulty was determined by the number of required resources, whether they should arrive in a specific order, and the speed of the resources needed to complete it. 

In Equation 1, variables (end - start) refer to how long it took the team to complete the event based on the start time of the event (when it became active) and the end time (when the last correct resource arrived). Also, (limit - start) is how long the event could be active until it was no longer available, and difficulty refers to the event's difficulty level. In Equation 2, Raw Score is the cumulative sum of the Event Scores, and Worst Score is the cumulative sum of the event scores assuming no event was completed, and Best Score is the cumulative sum of the event scores assuming all events were completed immediately. 

The team score also factors how quickly teams successfully responded to events, earning a higher possible score for responding rapidly and accurately to higher difficulty events than lower difficulty events. It was computed by subtracting the ratio of Raw Score and Worst Score (the Raw Score implied by the player taking no action) and multiplying by 100 to cast the result as a score between 0 and 100. By weighing events of all levels of severity, the Relative Score emphasizes performance across all events instead of the Raw Score, which was biased towards high difficulty events. 

Thus, team performance can be represented by the average of the teammates' relative scores (highest being best) and the cumulative sum of their raw scores (lowest is best). Table \ref{metrics_and_measurements} matches team metrics with particular measurements, with task performance and situational awareness being automatically computed by the NeoCITIES database.

\begin{table}[h]
    \centering
    \caption{Performance Metrics and Measurements}
    \label{metrics_and_measurements}
            \resizebox{\columnwidth}{!}{%
            \begin{tabular}{|c|c|}
                \hline
                 \textbf{Metric} & \textbf{Measurement} \\
                 \hline
                 Task Performance & Raw Score, Relative Score \\
                 \hline 
                 Situational Awareness & Synchronicity, Sequencing \\
                 \hline
                 Perception of Team Cognition & Survey Results \\
                 \hline
            \end{tabular}%
            }
\end{table}

\subsubsection{Team Situational Awareness}

Team situational awareness is assessed through two variables: sequencing and Synchronicity; both were based on past NeoCITIES equations \cite{hellar2010neocities} and were designed to assess the temporal situational awareness of teammates while performing the NeoCITIES task. These variables were explicitly developed to align with Endsley's model of situation awareness \cite{Endsley_Garland_2000}. Sequencing refers to the team's ability to apply resources in the correct order and indicate the team's ability to perform together through dynamic situations over time. For example, the correct resources applied to the event were divided by all resources sent and then multiplied by the number of resources required by the event for a maximum of three. On the other hand, Synchronicity refers to the degree to which the correct resources were applied to the event within a narrow timeframe. Synchronicity provides insight into the events that determine if the team did or did not act together within the pre-defined timeframe in the task \cite{Mohammed_2009_sym}. Synchronicity started at two and was subtracted by the time the resource first arrived minus when the last resource arrived. Subsequently, the maximum possible situational awareness score a team could achieve was five, while the lowest possible score was 0. High situational awareness in the NeoCITIES simulation is characterized by fluid and coordinated resource allocation, with each team member's resource arriving at multi-resource events very soon after one another (Synchronicity). High situational awareness teams would also send those resources to events in the correct order as dictated by the specific event, like the chemical truck spill (sequencing). Both these variables were automatically computed at the end of the session.

\subsubsection{Perception of Team Cognition}

Participant's perceptions of team cognition were measured by summing the results of three surveys across each human participant in the team, meaning only three of the four conditions gave results in this measure. Each survey utilized contained questions from Johnson and colleagues team-related knowledge measure, meant to collect the degree of shared knowledge among team members in five distinct emergent factors of shared mental models \cite{johnson2007measuring}. Their measure was utilized as it is based on eliciting confined and relevant constructs to shared mental models, which are the standard operationalization of team cognition \cite{mohammed2010metaphor}. Questions were taken from each factor of the measure that pertained to the specific facet of team cognition targeted. The final score each team could achieve ranged from 100, meaning a high perception of team cognition, to 0, meaning no perception of team cognition within the team. Answers were given on a five-point Likert scale anchored with one being "Disagree" to five being "Agree."

The first team cognition variable analyzed was team shared knowledge, a composite of perception-related sub-variables that range from perceived shared knowledge about the task to the perceived mutual understanding of individual preferences and communication tendencies. These sub-variables were elicited through a specific question in the teamwork principles survey and had a designated sub-variable identification. The block of questions for this variable includes questions such as "My teammate has a general knowledge of specific team tasks" and "My teammate strives to express his or her opinion."

The second element of team cognition analyzed was team environment, a composite of several sub-variables related to trust, perceived rewards tied to behavior, safety, and perceived constraints. All these items were once again measured through the survey and aggregated to generate the variable. The block of questions for this includes "There is an atmosphere of trust among my teammates" and "My team knows the environmental constraints when we perform various team tasks."

The last team cognition variable analyzed was emergent interaction, a composite of mutual role understanding, perceived shared information, perceived interaction level, perceived exchange effectiveness, flexibility, collaborative decision-making, informal communication, and listening. Once again, individual questions designated for each variable in the survey were pulled to measure this construct. The block of questions for this variable included "My team understands its roles and responsibilities" and "My teammates consistently demonstrate effective listening skills."

\section{Results}
Analyses were conducted on the outcome measures to address the research questions and are organized by dependent variable.

\subsection{Team Performance}
In regards to team performance two separate 4 (conditions) x 5 (iteration) split-plot Analysis of Variance (ANOVA) were conducted to determine whether the team composition (i.e. AI-only, human-AI-AI, human-human-AI, human-only) differed with respect to their performance and team situation awareness improvement over time (i.e. 5-iterations). The assumption of sphericity for repeated measures was not met and degrees of freedom were automatically adjusted using the Geisser-Greenhouse correction. According to the findings of the first split plot analysis, there was the main effect of condition \textit{F}(3, 38.6) = 509, \textit{p} $<$ .001, while there were no significant interaction effects of condition by iteration \textit{F}(12, 158) = 0.75, \textit{p} = .699 and no iteration main effect \textit{F}(4, 158) = 0.86, \textit{p} = .489.

According to the significant condition main effect, Tukey's Post-Hoc comparisons indicate that teams in the AI-only condition performed significantly better than the other three conditions (\textit{p} $<$ .001). While teams in human-human-AI and human-AI-AI conditions performed equally (\textit{p} = .948), they performed significantly better than the human-only condition (\textit{p} $<$ .001; see Figure 4). Overall, these results show that their overall performance decreased when a human team member was included in a team. A reason for this outcome can be understood in the context of the strengths of the contemporary RL models used to develop the NeoCITIES agents. RL agents have recently achieved superhuman performance across a variety of domains \cite{foerster2018learning}.

\begin{figure} [h!]
  \label{team_score_means_plot}
  \centering
    \includegraphics[width=0.35\textwidth]{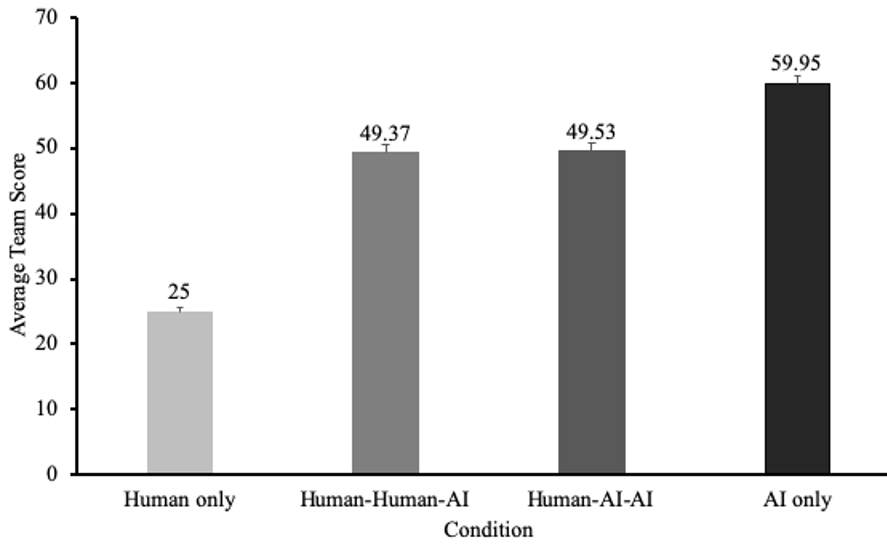}
    \caption{Average Team Performance Score across the Conditions (Vertical lines provide +SE)}
\end{figure}

\subsection{Team Situation Awareness}
Next, a 3x5 split-plot analysis was conducted and tests similar to team performance (score) were ran for team situational awareness. Just as the previous repeated measures ANOVA, the assumption of sphericity was not met, and the degrees of freedom were automatically adjusted to correct for the potential of false positive results. Just as with team score, there was the main effect of condition \textit{F}(3, 40.1) = 239, \textit{p} $<$ .001, while there was no significant interaction effect of condition by iteration \textit{F}(12, 159) = 1.49, \textit{p} = .132 and no interaction main effect \textit{F}(4, 159) = 0.79, \textit{p} = .531.

According to the significant condition main effect, the human-only teams significantly under-performed (\textit{p} $<$ .001), which makes sense given how team situational awareness is highly related to team score (higher team situational awareness necessarily means that many resources reached events to complete a given task). All three other conditions did not differ on team situation awareness: AI-only vs human-AI-AI (\textit{p} = .37), AI-only vs human-human-AI (\textit{p} = .69), and human-AI-AI vs human-human-AI (\textit{p} = .63; see Figure 5). Overall, these findings make sense given the large gap in team situational awareness between human-only teams and the other teams and the far lesser gap between HATs and AI-only teams.

\begin{figure} [h!]
  \label{situational_awareness_means_plot}
  \centering
    \includegraphics[width=0.35\textwidth]{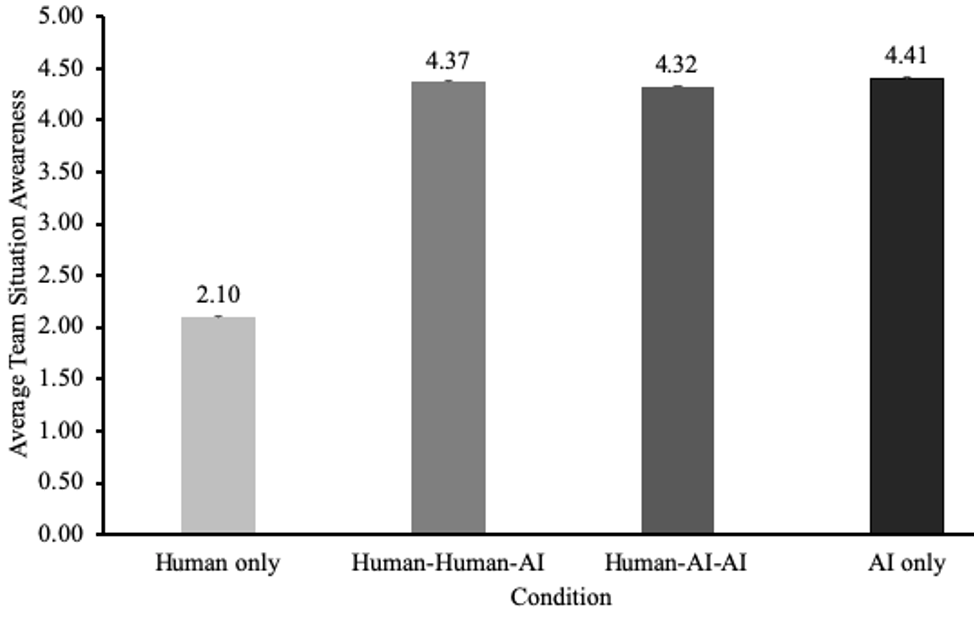}
     \caption{Average Team Situation Awareness across the Conditions (Vertical lines provide +SE)}
\end{figure}

\subsection{Team Cognition Survey Analysis}
Here the focus is shifted exclusively toward the teams that included human participants (thereby excluding the AI-only teams) to not only examine the existence of perceived team cognition within these HATs but to analyze the ability to utilize perceived team cognition as a predictor of teamwork for HATs. This analysis was accomplished through a linear regression model that relates perceived team cognition variables to team score and team situational awareness.

All three variables were combined into one regression model for both team score (team\_score $\sim$ team\_knowledge $+$ team\_environment $+$ team\_interaction) and team situational awareness (team\_situational\_awareness $\sim$ team\_knowledge $+$ team\_environment $+$ team\_interaction). These three variables (team knowledge, team environment, and team interaction), once combined, serve as proxies for team cognition as defined in the current study's methodology. The results are displayed in Table \ref{lr_team_score}. 

\begin{table}[h!]
    \centering
    \caption{Team Cognition Linear Regression for Team Score}
    \label{lr_team_score}
        \begin{tabular}{|c|c|c|c|}
             \hline
             Variable & Coefficient & Std Error & P-value \\
             \hline
             \multicolumn{4}{|c|}{Score} \\
             \hline
             Team Knowledge & -.42 & 2.75 & .07 \\
             \hline 
             Team Environment & -1.09 & .54 & .05 \\
             \hline 
             Team Interaction & -1.08 & .43 & .02 \\
             \hline 
             \multicolumn{4}{|c|}{Team Situational Awareness} \\
             \hline
             Team Knowledge & -.05 & .02 & .02 \\ 
             \hline  
             Team Environment & -.05 & .05 & .29 \\ 
             \hline 
             Team Interaction & -.11 & .04 & .01 \\
             \hline
        \end{tabular}
\end{table}

The "score" model had a residual standard error of 3.089 on 26 degrees of freedom. Looking at the R-squared value, the perceived team cognition variables collectively explain 93.99\% of the team scores variance, \textit{F}(3, 26) = 135.4, which was significant (\textit{p} = $<$ .001). Pursuant to these results, it becomes possible to regularize the coefficients and produce beta coefficients that can be more effectively interpreted.

Considering each perceived team cognition variable's statistical significance, only team interaction has a meaningful influence on team score. Specifically, the team score is expected to decrease by 0.46 for every standard deviation increase in team interaction. On the other hand, the team score is expected to decrease by 0.26 for every standard deviation increase in the team environment; however, these findings did not reach significance (\textit{p} = .054). These results need to be understood in context: the model's intercept is 83.67. Thus, across conditions, it seeks to predict performance losses between human-only and HATs. Therefore, the results suggest that even though perceived team cognition is more prevalent as more humans become part of the team, performance decreases with perceived team cognition because it emerges more intensively among the humans in the underperforming human-only teams than humans and AI. 

\begin{figure} [h!]
  \label{perc of team cog}
  \centering
    \includegraphics[width=0.35\textwidth]{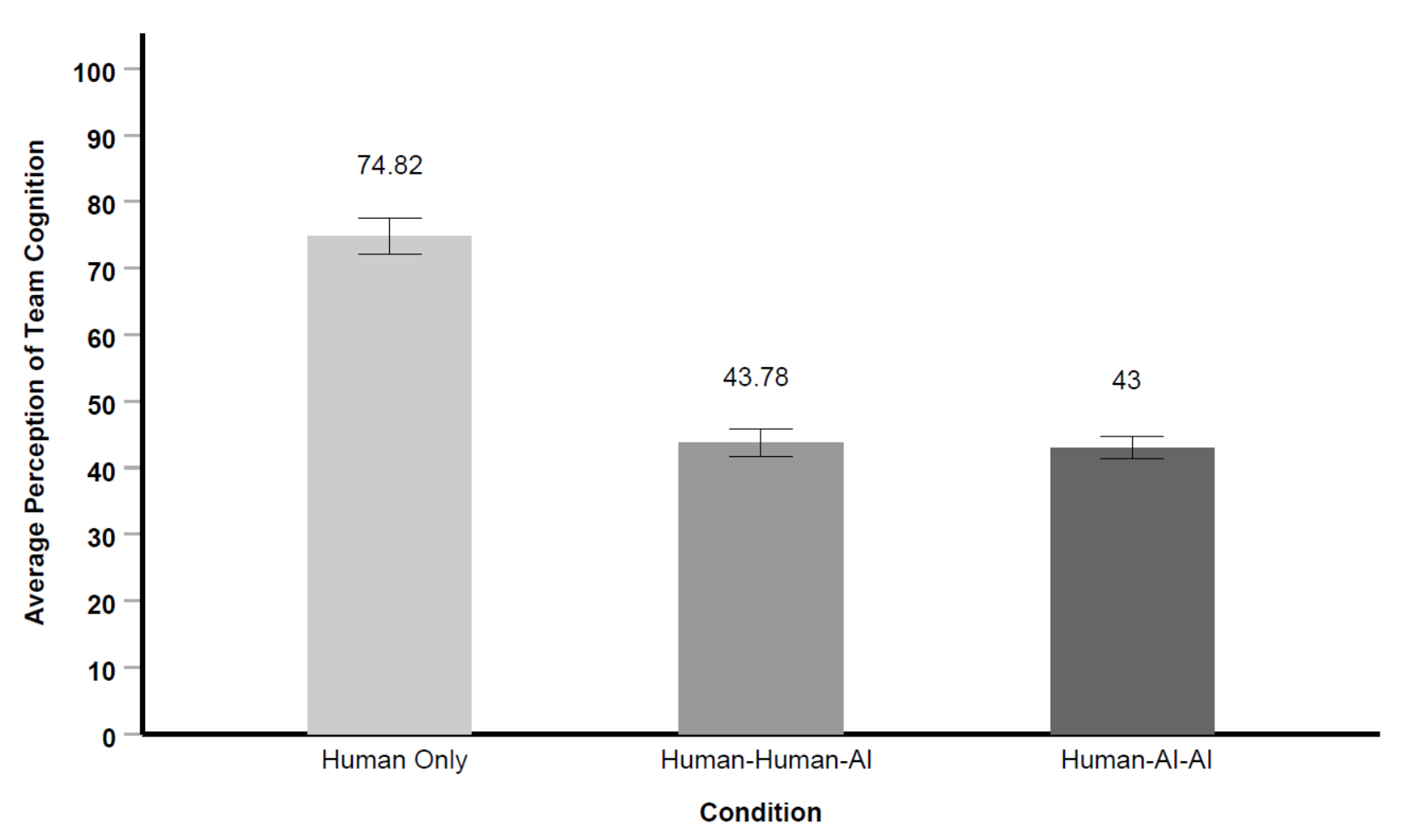}
     \caption{Average Perception of Team Cognition across Conditions Involving Humans}
\end{figure}

On the other hand, the "team situational awareness" model had a residual standard error of 0.29 on 26 degrees of freedom. Multiple R-squared shows that perceived team cognition explained most of the variance in team situational awareness (93.91\%). As expected, the F-statistic, \textit{F}(3, 260) = 133.7, is closely related to that of the "score" model and is significant. \textit{p} = $<$ .001. Again likely due to the fact that both performance variables were highly related because of the nature of the simulation. Again, given the strength of the model, the coefficients were regularized into beta coefficients. 

Team situational awareness is expected to decrease by 0.36 for every standard deviation increase in team knowledge (\textit{p} = .02). On the other hand, the team environment was not statistically significant. Lastly, team situational awareness is expected to decrease by 0.49 for every standard deviation increase in team interaction (\textit{p} = 0.01). Once again, the model's intercept is 7.33; hence it is oriented towards understanding performance drop-offs between HATs and human-only teams, and it does so very effectively.

Finally, as seen in Figure 6, the perception of team cognition suffers significantly with the introduction of an artificial agent on a team, falling from 74.82 in the all human condition to 43.78 and 43 in the human-human-AI and human-AI-AI conditions respectively. The perception of team cognition within teams involving humans fell sharply (by 31.04) with the inclusion of a single AI, but the inclusion of another AI within the team had a relatively negligible effect, only falling by 0.78.

\section{Discussion}
The study enables us to draw inferences from several different perspectives. The different experimental conditions allow us to examine different compositions of HATs, determining how interacting with AI changes human players' ability to perform and adapt to the environment and the extent to which different compositions affect team characteristics and dynamics. 

\subsection{Perceived Team Cognition}
In terms of answering RQ1, which asked how team characteristics and dynamics are similar and or different across different team compositions, these results run counter to expectations based on extending what the literature has identified as critical aspects of human teamwork. Indeed, HATs outperformed human-only teams despite lower levels of perceived team cognition.

Specifically, as a predictor of team performance, the data indicates that higher levels of team interaction results in lower team scores. Here RQ2, which asked how these results can help inform future HATs, is addressed. One interpretation of this result is that introducing an AI into the NeoCITIES team often induces the human players to communicate differently to compensate for the agent's inability to coordinate directly. However, performance is still higher even in teams with only one human, which, in the context of AI-only teams outperforming all other teams, suggests that humans can achieve higher team scores following the AI's lead (depending on task and context). Still, concerning team score, the other perceived team cognition variables were not significant, limiting the inferences that may be drawn about the relationship between perceived team cognition and team score alone among teams that included humans. 

Continuing to respond to RQ2, analyzing the perception of team cognition in each team gives us insight into how team composition and the introduction of an artificial agent affects this team characteristic. The perception of team cognition in all human teams was much higher than in teams involving an AI, which is in line with prior literature. It remains interesting that some perception of team cognition within HATs existed, each with a score of at least 42 from the possible 100. This result is not surprising when one considers the prior literature but is critical to consider as the research community and developers alike make attempts to close the gap in team cognition between HATs and traditional human-human teams. There is a great deal of work that remains to be done to bring these teams' shared understanding closer to one another, as it is apparent that the introduction of an AI negatively affected perceived team cognition as a teammate. While performance was better in the HATs than the human-human teams, despite the lower levels of perceived team cognition, this can likely be attributed to expert level AI agents created through RL \cite{foerster2018learning}. However, there was still a considerable amount of team cognition present on the HATs at approximately a score of 43; what is even more interesting is that the human-AI-AI condition had nearly the same level of perceived team cognition as the human-human-AI condition. This result is surprising as it would be expected that the lack of another human on the team would severely undercut team cognition, but this was not the case in the human-AI-AI condition. Finally, this result could open the door for several different potential tweaks in agent design, task environment, interface, or other changes for HATs to further develop and or accommodate perceived team cognition. An example of which could be enhanced displays that engage teams in the task spatially to showcase teammate status and action tendencies over time.

\subsubsection{Team Situational Awareness}
Answering RQ1 in light of team situational awareness provides a deeper analysis of perceived team cognition within these teams. These results show that team knowledge and team interaction negatively influenced team situational awareness, except for team environment. Specifically, team interaction had twice the impact as team knowledge on team situational awareness. Moreover, in regards to RQ2, one way to consider this result is that the presence of an AI teammate who cannot communicate forces the human players to redirect their efforts and more effectively coordinate themselves through NeoCITIES' map, thereby reacting more rapidly as they try to synchronize with the autonomous agent. This finding provides further evidence for utilizing agents in tasks that lend themselves to being displayed in their entirety on an interface to convey the task situation and teammate actions, helping team members develop team cognition through heightened situational awareness and team member tendencies.

\subsection{Performance Data}
Beyond the survey, the performance data is the last component of RQ1. Surprisingly, the HATs did not outperform AI-only teams. This finding is surprising as there is precedence for HATs outperforming both human-only teams as well as AI-only teams \cite{wang2016deep}. The AI-only teams substantially outperformed all other team types across all performance metrics, whereas human-only teams underperformed all other team types in performance metrics. The results also show that the HATs performed closer to AI-only teams (17\% lower than AI-only vs. doubling human-only). As far as the relatively surprising case of HATs not outperforming AI teams, it seems the results more closely match those of DeepMind and OpenAI, where over time, an RL agent achieves superhuman performance through strategies, tactics, and responses that run counter to human intuition and is noted in the literature \cite{foerster2018learning}.

Given how this is the first time NeoCITIES has been used to study human-AI teamwork, many of the prior results outlined in the literature are thus not as applicable because they only speak to human teams instead of HATs. However, one way to frame this finding in response to RQ2 is by focusing on RL's unique dynamics. Whereas prior attempts at studying HATs have suffered from technical limitations (automated as opposed to autonomous agents, Wizard-of-Oz simulations of agents by humans), RL may enable the development of agents that eventually get so strong at the task that they can operate almost independently. Alternatively, they may operate more effectively as part of a multi-agent system with only AI instead of one where humans are also involved. These AI agents acting with superhuman ability are part of why the mixed teams resulted in higher performance scores than the all-human teams despite the lower levels of more traditional performance predictors like team cognition. The agents' superior ability and lack of communication potentially force the human teammate to adapt their typical teaming behavior, which should be accounted for when considering the level of ability with which an artificial agent will operate.

\subsection{Effects of Team Composition}
Essentially, the composition of teams created several differences and similarities between the conditions. Perceived team cognition showcased how the difference between the two conditions of HATs was relatively negligible, indicating that the addition of more AI had a less drastic effect on perceived team cognition than would be expected based on its drop with the addition of a single AI. Regarding the performance between teams, it was surprising to find that both types of HATs performed very similarly. While human-human-AI teams scored on average 0.32\% lower for the team score, their team situational awareness was, on average, 1.1\% higher than human-AI-AI teams. These patterns are important because they show that it is not necessarily the case that human-AI-AI teams succeed because of the higher number of agents. Furthermore, the gap between HATs and AI-only teams is far narrower for team situational awareness (1\% difference) than for scores (17\%). This result suggests that although the coordination behavior between these teams is very different (for example, human-AI-AI teams cannot rely on communication), both AI and humans can adapt to the circumstances to retain high-performance levels and team situation awareness. Although they succeed for different reasons, they succeed in much the same way at a complex task such as emergency response management.

\section{Limitations}

The current experiment contains a few limitations to keep in mind when interpreting and applying these results. The first limitation to keep in mind is the NeoCITIES task nature, which is emergency response management. Despite the long history of the NeoCITIES platform in team research \cite{hellar2010neocities}, teams may perform differently in a more relaxed context (not emergency response management). Finally, the fact that the reinforcement learning AI model was not trained with humans can be a limiting factor when considering these results, as past research has shown improvements in HAT performance when utilizing this training strategy \cite{Carroll_2019}. However, that same research did showcase that the model trained with humans could not take advantage of the adaptations to team behaviors humans make over time, which is necessary to support team cognition.

\section{Conclusion}
In essence, these results suggest that the best way to understand HATs' complex dynamics is not by attempting to replicate human teamwork dynamics through the design of human-like agents but by leveraging differences in AI with tasks and environments that take advantage of those differences. Perceived team cognition was not a predictor of performance, which is contrary to the prior literature. The teams' composition also showcased drastic differences between human-human team performance and human-AI performance, which is paralleled by their differences in perceived team cognition. However, there were few differences in the two human-AI conditions regarding performance, team situation awareness, and perceived team cognition. This result indicates that mixed teams are not severely affected by the inclusion of additional agents; however, behavior among the humans across the different conditions changes based on the presence and number of agents on the team.

\section{Appendix}
\subsection{NeoCITIES Scenario Information}
The NeoCITIES scenarios begin immediately following the group training with all players. The scenario is presented in training as an emergency response management task where the teammates' responsibility is to respond to several emergency events occurring within a fictional city. Each of the emergency events plaguing the fictional city are presented with a description to give the scenario additional depth; for example, the description of the Football Weekend Briefing shown to players is as follows: "Unit(s) needed to deliver a talk for the upcoming football weekend on how to identify potential bombs and bomb-making materials." Another example of an emergency event description is: "Caller reports falling asleep with a lit cigarette has led to a fire in his apartment. Resident is unable to contain the fire in the room and needs assistance." These descriptions of emergency events are shown in the players' task cards, where they gather the information needed to respond to the events. Consequentially, the event descriptions go hand in hand with the premise of the NeoCITIES simulation given to the players in their training phase to create an encompassing and detailed scenario.

\bibliographystyle{IEEEtran}
\bibliography{cites.bib}

\begin{IEEEbiography}
    [{\includegraphics[width=1in,height=1.25in,clip,keepaspectratio]{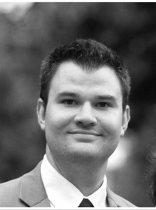}}]{Nathan McNeese}
\textbf{Dr. Nathan J. McNeese} (M’18) is an Assistant Professor and Director of the Team Research Analytics in Computational Environments Research Group in the School of Computing at Clemson University. Dr. McNeese received his Ph.D. in Information Sciences \& Technology with a focus on Team Decision Making, and Cognition from The Pennsylvania State University in Fall 2014. His current research interests include the study of human-AI teaming, and the development/design of human-centered collaborative systems.
\end{IEEEbiography}

\begin{IEEEbiography}
    [{\includegraphics[width=1in,height=1.25in,clip,keepaspectratio]{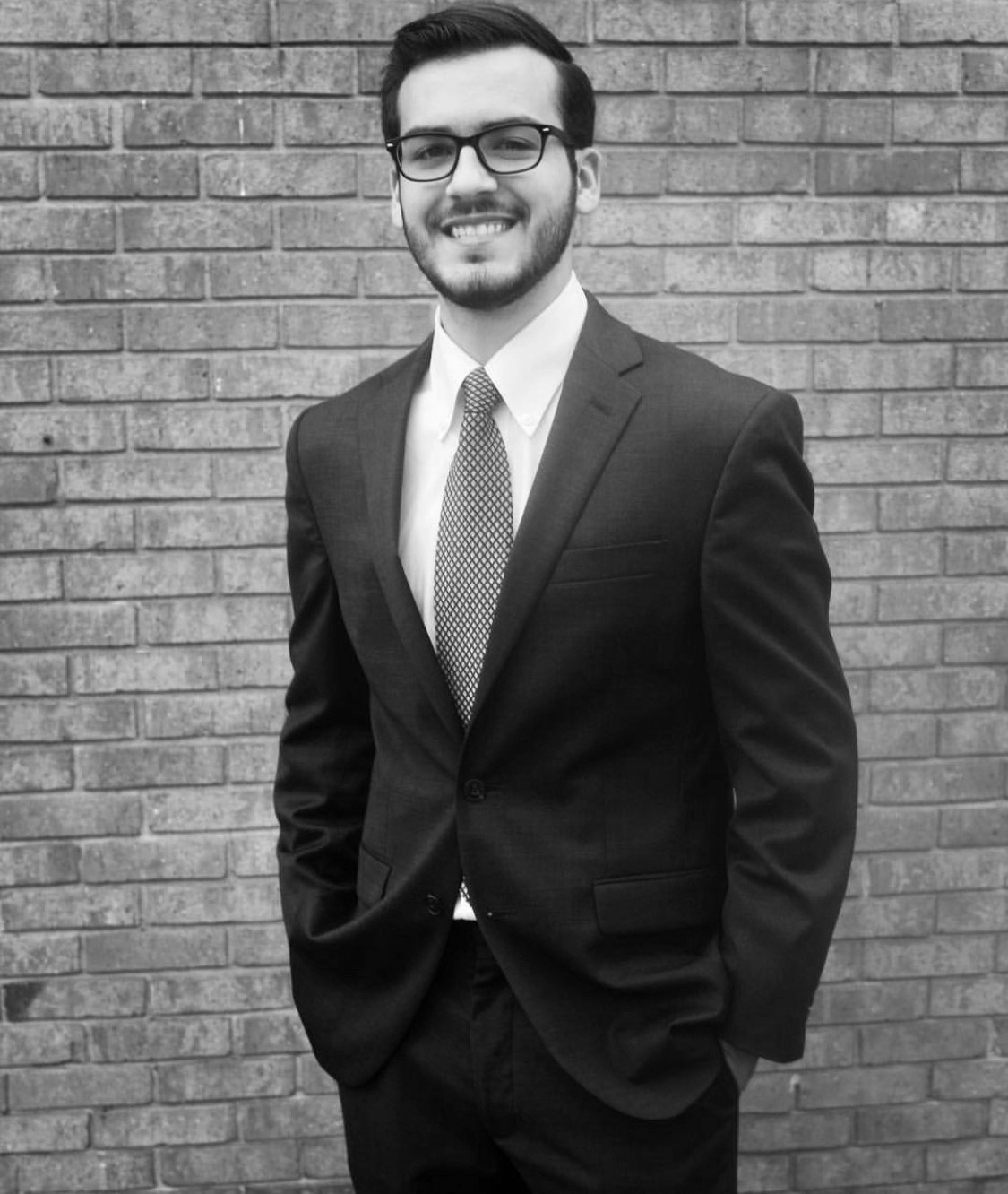}}]{Beau Schelble}
\textbf{Mr. Beau Schelble} is a Ph.D. student in Human-Centered Computing at Clemson University. He is a member of the Team Research Analytics in Computational Environments Research Group within the Clemson University School of Computing. His current research interests include team cognition within human-AI teams, human-centered AI, multi-agent teaming.
\end{IEEEbiography}

\begin{IEEEbiography}
    [{\includegraphics[width=1in,height=1.25in,clip,keepaspectratio]{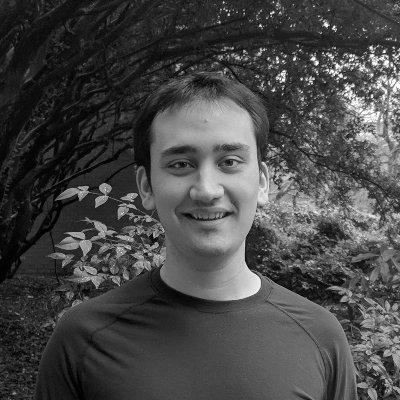}}]{Lorenzo Barberis Canonico}
\textbf{Dr. Lorenzo Barberis Canonico} is a Bioinformatics Researcher at Stanford University. He received his Ph.D. in Human-Centered Computing from Clemson University in 2019. His current research focus revolves around the intersections of game theory, machine learning, and collective intelligence. His primary areas of research are prediction markets, multi-agent systems, reinforcement learning and crowdsourcing applications.
\end{IEEEbiography}

\begin{IEEEbiography}
    [{\includegraphics[width=1in,height=1.25in,clip,keepaspectratio]{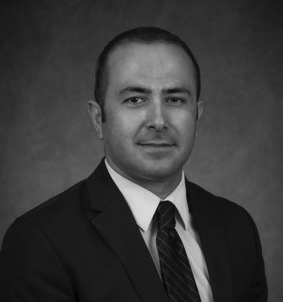}}]{Mustafa Demir}
\textbf{Dr. Mustafa Demir} (M’15) is currently a Senior Quantitative Research Scientist \& faculty associate working at Ira. A. Fulton Schools of Engineering at Arizona State University. Dr. Demir received his Ph.D. in Simulation, Modelling and Applied Cognitive Science with a focus on team coordination dynamics in Human-Autonomy Teaming from Arizona State University in Spring 2017. His current research interests are human-autonomy teaming, team cognition, dynamical systems theory and advanced statistical modelling. 
\end{IEEEbiography}

\end{document}